\documentclass[twocolumn,aps,showpacs,showkeys,preprintnumbers,superscriptaddress,amsmath,amssymb,floatfix]{revtex4}

\usepackage{graphicx}
\usepackage{dcolumn}
\usepackage{bm}

\usepackage{color}

\begin{document}

\definecolor{red}{rgb}{.9,.1,.1}
\definecolor{green}{rgb}{.1,.9,.1}
\definecolor{blue}{rgb}{.1,.1,.9}
\newcommand{\red}[1]{{\color{red} #1}}
\newcommand{\green}[1]{{\color{green} #1}}
\newcommand{\blue}[1]{{\color{blue} #1}}

\preprint{APS/valid preprint number appear here}

\title{Astrophysical S-factor of the $^{3}$He($\alpha$,$\gamma$)$^{7}$Be reaction measured at low energy via prompt and delayed $\gamma$ detection }

\author{F. Confortola}
\affiliation{Universit\`a degli Studi Genova \& INFN Genova, Via
Dodecaneso 33, 16146 Genova, Italy}

\author{D. Bemmerer}
\altaffiliation[Present address: ]{Forschungszentrum
Dresden-Rossendorf, Dresden, Germany} \affiliation{INFN Sezione di
Padova, Via Marzolo 8, 35131 Padova, Italy}

\author{H. Costantini}
\affiliation{Universit\`a degli Studi Genova \& INFN Genova, Via
Dodecaneso 33, 16146 Genova, Italy}

\author{A. Formicola}
\affiliation{INFN, Laboratori Nazionali del Gran Sasso, S.S. 17bis
km 18.890, Assergi, L'Aquila, Italy}

\author{Gy. Gy\"urky}
\affiliation{ATOMKI, Debrecen, Hungary}

\author{P. Bezzon}
\affiliation{INFN, Laboratori Nazionali di Legnaro, Padova, Italy}

\author{R. Bonetti}
\affiliation{Istituto di Fisica Generale Applicata, Universit\`a
di Milano \& INFN Milano, Milano, Italy}

\author{C. Broggini}\thanks{Corresponding author: \emph{E-mail address}: broggini@pd.infn.it}
\affiliation{INFN Sezione di Padova, Via Marzolo 8, 35131 Padova,
Italy}

\author{P. Corvisiero}
\affiliation{Universit\`a degli Studi Genova \& INFN Genova, Via
Dodecaneso 33, 16146 Genova, Italy}

\author{Z. Elekes}
\affiliation{ATOMKI, Debrecen, Hungary}

\author{Zs. F\"ul\"op}
\affiliation{ATOMKI, Debrecen, Hungary}

\author{G. Gervino}
\affiliation{Dipartimento di Fisica Sperimentale, Universit\`a di
Torino \& INFN Torino, Torino, Italy}

\author{A. Guglielmetti}
\affiliation{Istituto di Fisica Generale Applicata, Universit\`a
di Milano \& INFN Milano, Milano, Italy}

\author{C. Gustavino}
\affiliation{INFN, Laboratori Nazionali del Gran Sasso, S.S. 17bis
km 18.890, Assergi, L'Aquila, Italy}

\author{G. Imbriani}
\affiliation{INAF, Osservatorio Astronomico di Collurania, Teramo,
Italy}

\author{M. Junker}
\affiliation{INFN, Laboratori Nazionali del Gran Sasso, S.S. 17bis
km 18.890, Assergi, L'Aquila, Italy}

\author{M. Laubenstein}
\affiliation{INFN, Laboratori Nazionali del Gran Sasso, S.S. 17bis
km 18.890, Assergi, L'Aquila, Italy}

\author{A. Lemut}
\affiliation{Universit\`a degli Studi Genova \& INFN Genova, Via
Dodecaneso 33, 16146 Genova, Italy}

\author{B. Limata}
\affiliation{Dipartimento di Scienze Fisiche, Universit\`a
``Federico II'' \& INFN Napoli, Napoli, Italy}

\author{V. Lozza}
\affiliation{INFN Sezione di Padova, Via Marzolo 8, 35131 Padova,
Italy}

\author{M. Marta}
\affiliation{Istituto di Fisica Generale Applicata, Universit\`a
di Milano \& INFN Milano, Milano, Italy}

\author{R. Menegazzo}
\affiliation{INFN Sezione di Padova, Via Marzolo 8, 35131 Padova,
Italy}

\author{P. Prati}
\affiliation{Universit\`a degli Studi Genova \& INFN Genova, Via
Dodecaneso 33, 16146 Genova, Italy}

\author{V. Roca}
\affiliation{Dipartimento di Scienze Fisiche, Universit\`a
``Federico II'' \& INFN Napoli, Napoli, Italy}

\author{C. Rolfs}
\affiliation{Institut f\"ur Experimentalphysik III,
Ruhr-Universit\"at Bochum, Bochum, Germany}

\author{C. Rossi Alvarez}
\affiliation{INFN Sezione di Padova, Via Marzolo 8, 35131 Padova,
Italy}

\author{E. Somorjai}
\affiliation{ATOMKI, Debrecen, Hungary}

\author{O. Straniero}
\affiliation{INAF, Osservatorio Astronomico di Collurania, Teramo,
Italy}

\author{F. Strieder}
\affiliation{Institut f\"ur Experimentalphysik III,
Ruhr-Universit\"at Bochum, Bochum, Germany}

\author{F. Terrasi}
\affiliation{Dipartimento di Scienze Ambientali, Seconda
Universit\`a di Napoli, Caserta \& INFN Napoli, Napoli, Italy}

\author{H.P. Trautvetter}
\affiliation{Institut f\"ur Experimentalphysik III,
Ruhr-Universit\"at Bochum, Bochum, Germany}

\collaboration{The LUNA Collaboration} \noaffiliation

\date{\today}

\begin{abstract}
Solar neutrino fluxes depend both on astrophysical and on nuclear
physics inputs, namely on the cross sections of the reactions
responsible for neutrino production inside the Solar core. While
the flux of solar $^8$B neutrinos has been recently measured at
Superkamiokande with a 3.5\% uncertainty and a precise measurement
of $^7$Be neutrino flux is foreseen in the next future, the
predicted fluxes are still affected by larger errors. The largest
nuclear physics uncertainty to determine the fluxes of $^8$B and
$^7$Be neutrinos comes from the $^3$He($\alpha$,$\gamma$)$^7$Be
reaction. The uncertainty on its S-factor is due to an average
discrepancy in results obtained using two different experimental
approaches: the detection of the delayed $\gamma$ rays from $^7$Be
decay and the measurement of the prompt $\gamma$ emission. Here we
report on a new high precision experiment performed with both
techniques at the same time. Thanks to the low background
conditions of the Gran Sasso LUNA accelerator facility, the cross
section has been measured at E$_{cm}$ = 170, 106 and 93 keV, the
latter being the lowest interaction energy ever reached.
 The S-factors from the two methods do not show any
discrepancy within the experimental errors. An extrapolated S(0)=
0.560$\pm$0.017 keV barn is obtained. Moreover, branching ratios
between the two prompt $\gamma$-transitions have been measured
with 5-8\% accuracy.

\end{abstract}


\pacs{25.55.-e, 26.20.+f, 26.65.+t}
\keywords{$^{3}$He($\alpha$,$\gamma$)$^{7}$Be, p-p chain, direct
measurement, underground accelerator}

\maketitle

Forty years ago, John Bahcall and Raymond Davis started to explore
the solar interior by studying the neutrinos emitted by the Sun
\cite{bahcall64}. The results of the first neutrino detection
experiment \cite{davis64} originated the so called solar neutrino
puzzle, consisting in a deficit of measured neutrinos with respect
to the theoretical predictions of the Standard Solar Model (SSM).
After thirty years of experiments, SNO and Kamland
\cite{sno,kamland} observed neutrino oscillations and proved that
the missing solar electron neutrinos actually change their flavour
during the travel to the Earth. This closed the neutrino puzzle.
Therefore, the high precision measurement of $^8$B neutrino flux
\cite{SKK}, together with the foreseen measurement of $^7$Be
neutrinos \cite{borexino}, can now be used to understand physical
and chemical properties of the Sun, provided that nuclear reaction
cross sections are known with similar accuracy \cite{fiorentini}.
 The $^3$He($\alpha$,$\gamma$)$^7$Be reaction is the onset of the $^7$Be and $^8$B branches of the proton-proton (p-p)
 chain of hydrogen burning. The 9\% error \cite{adelberger} on its cross section is presently the
 main nuclear physics uncertainty on the prediction of $^7$Be and $^8$B neutrino fluxes \cite{bahcall04}.

At stellar energies the $^3$He($\alpha$,$\gamma$)$^7$Be cross
section $\sigma$(E) drops exponentially with the energy and can be
parametrized as:
\begin{equation}
    \sigma(E) = \frac{S(E)}{E}e^{-2\pi\eta(E)}
    \label{csdef}
\end{equation}
where  S(E) is the astrophysical factor, $\eta$ is the Sommerfeld
parameter \cite{rolfs}, and E is the center of mass energy.

$^3$He($\alpha$,$\gamma$)$^7$Be is a radiative capture reaction
(Q-value: 1.586 MeV) into the first excited state (E$_x$=429 keV)
and the ground state of $^7$Be that subsequently decays by
electron capture into $^7$Li with a terrestrial half life of
53.22$\pm$0.06 days \cite{tilley}.

In the last forty years the $^3$He($\alpha$,$\gamma$)$^7$Be
reaction has been measured using two techniques. In the first
approach direct $\alpha$-capture $\gamma$-rays were detected
(prompt $\gamma$ method)
\cite{osborne,holmgren,parker,nagatani,krawinkel, alexander,
hilgemeier}, while, in the second, the delayed $^7$Be-decay
$\gamma$ rays were counted (activation method)
\cite{osborne,robertson,volk,singh}. Previous activation results
are, on the average, 13\% higher than prompt $\gamma$ data and
this is the origin of the large uncertainty quoted
 on the reaction cross section \cite{adelberger}.
 Up to now, no explanation has been obtained for this discrepancy
 that could be due either to systematic experimental errors
 (angular distribution, branching ratio effects, parasitic reactions producing $^7$Be) or to the existence of a non radiative capture (E0 monopole) \cite{E02}.
Recently, the discrepancy on the extrapolated S(0) has been
reduced by an activation study at 420$<$E$<$950 keV \cite{singh}
to 9\%. High accuracy (4\%) activation data were obtained also at
LUNA, at center of mass energies down to 106 keV
\cite{bemmerer2,giuri}. Nevertheless, high accuracy prompt gamma
data are also needed to verify the claimed discrepancy.

 We have performed a new high accuracy measurement using simultaneously
 prompt and activation methods with the same experimental setup.
 The experiment has been carried out using the underground LUNA 400 kV
 accelerator
\cite{acceleratore} at the Gran Sasso National Laboratory (LNGS).

\begin{figure}
    \includegraphics[width=0.55\textwidth]{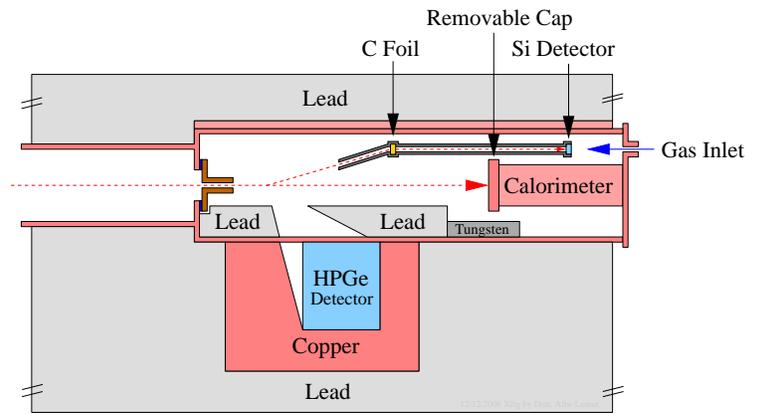}
    \caption{Schematic view of the interaction chamber with the position of the HPGe detector and of the 100 $\mu$m silicon detector used for $^3$He
    density monitoring. The distance between the entrance  collimator and the calorimeter is 35 cm. The thickness of the internal collimator is 3 cm for the Lead part and 1.6 cm for
    the Tungsten part.}
    \label{chamber}
\end{figure}

Three couples of cross section values  have been measured (prompt
$\gamma$ and activation) at E$_\alpha$=220, 250 and 400 keV.
 A sketch of the interaction chamber is given in Fig.
 \ref{chamber}.
The $\alpha$ beam enters the $^3$He extended windowless gas target
\cite{giuri} through a 7 mm diameter collimator and is stopped on
a detachable copper disk that serves as the primary catcher for
the produced $^7$Be and as the hot side of a calorimeter
\cite{casellatec}. The latter measures the beam intensity (about
250 $\mu$A) with an accuracy of 1.5\%. The high beam current
decreases the $^3$He density along the beam path \cite{goerres}:
this effect has been monitored with a silicon detector by double
Rutherford scattering providing an accuracy of 1.3\% on the gas
density determination \cite{marta}. The same detector is also used
to measure the gas contamination (mainly N$_2$) that has remained
below (2.7 $\pm$ 0.3)\%.

Prompt $\gamma$ rays are counted  with a 135\% ultra
low-background HPGe detector shielded with 5 cm of OFHC copper and
25 cm of lead. The detector and the shield are enclosed in a
sealed plastic box flushed with dry N$_2$ to reduce $^{222}$Rn
background.
 Thanks to the underground environment where cosmic muons are
 strongly reduced \cite{arpesella}, the shielding suppression factor is of five orders of magnitude for $\gamma$ rays below 2 MeV.
A lead collimator is positioned inside the target chamber to
collect mostly $\gamma$ rays emitted at 55$^o$. At this angle the
contribution of the second Legendre polynomial in the angular
distribution expression, vanishes.
 Therefore the inner collimator reduces the systematic
error due to prompt $\gamma$ angular distribution uncertainties
and also shields the detector from possible beam-induced radiation
coming from the entrance collimator and the calorimeter cap. The
effective target length seen by the HPGe detector is approximately
12 cm, corresponding to an energy loss $\Delta$E = 3 keV at
P$_{target}$=0.7 mbar (value used in all the runs) and
E$_{\alpha}$=400 keV.

\begin{figure}
    \includegraphics[width=0.5\textwidth]{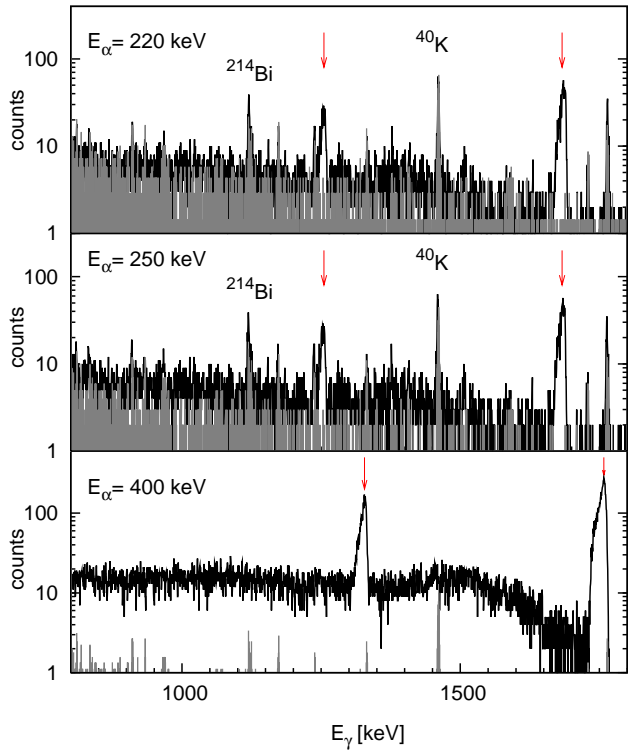}
    \caption{$\gamma$-ray spectrum at E$_\alpha$ = 220, 250 and 400 keV compared  with natural laboratory
    background (in grey) normalized to beam-measurement live times (respectively: 31.2 days, 21.3 days and
    4.8 days). Arrows indicate the primary transition peaks to the first excited state and to the ground state.}
    \label{spectra}
\end{figure}

The photopeak detection efficiency is determined by a Monte Carlo
code \cite{MC} calibrated with $^{60}$Co and $^{137}$Cs
radioactive point-like sources moved along the beam path. The
Monte Carlo reproduces the experimental efficiency within the
source activity uncertainties (1.5\%).
 The spectra collected at E$_{\alpha}$= 220, 250 and 400 keV, with a total charge of 637, 407 and 113 C respectively,
 are shown
 in  Fig. \ref{spectra} together with laboratory background. Beam induced gamma-ray background has been measured
with $^4$He gas in the target at E$_{\alpha}$=
 400 keV: no
 difference with laboratory background has  been observed.
 In the data analysis only the two primary transitions at E$_{\gamma}$=Q$+$E$_{cm}$ and
E$_{\gamma}$=Q$+$E$_{cm}$$-$429 keV, have been considered.
Theoretical angular distribution functions are calculated by
\cite{parker-teo} down to 210 keV interaction energy. A linear
extrapolation of the curves of \cite{parker-teo} has been done and
the coefficients of the Legendre polynomials adopted in the
detection efficiency calculation are: a$_1$= - 0.05 and a$_1$ = 0
for the transition to the ground and to the first excited state,
respectively and a$_2$= - 0.1 for both transitions. These values
are in agreement with recent theoretical predictions \cite{kim}.
With the Monte Carlo code, we have conservatively varied 100\%
both a$_1$ and a$_2$ coefficients obtaining a global 2.5\%
variation of the detection efficiency. This value has been assumed
as a systematic uncertainty and turns out to be the major
contribution to the error budget of the prompt $\gamma$ method.
 The in-beam runs provide accurate branching
ratios between the two transitions $\sigma$(DC$\to$
429)/$\sigma$(DC$\to$0): 0.417 $\pm$0.020, 0.415$\pm$0.029 and
0.38$\pm$0.03 at E$_{\alpha}$ = 400, 250 and 220 keV,
respectively. Our values are consistent with, but more precise
than, latest branching ratio measurements \cite{osborne,krawinkel}
and are in agreeement with theoretical calculations
\cite{kajino1}.

 During the runs in which prompt $\gamma$-rays are detected,
  the $^7$Be nuclei produced inside the gas target
get implanted into the removable calorimeter cap. After each run,
the cap is dismounted and moved to LNGS underground low-activity
counting facility \cite{arpesella}. Details and accuracy of the
activation method at LUNA are discussed elsewhere
\cite{bemmerer2,giuri}. Since we have simultaneously used the same
beam and target for both methods, some systematic uncertainties
(beam intensity, target density and purity) cancel out in the
comparison between the two techniques.

\begin{figure*}
    \includegraphics[width=1\textwidth]{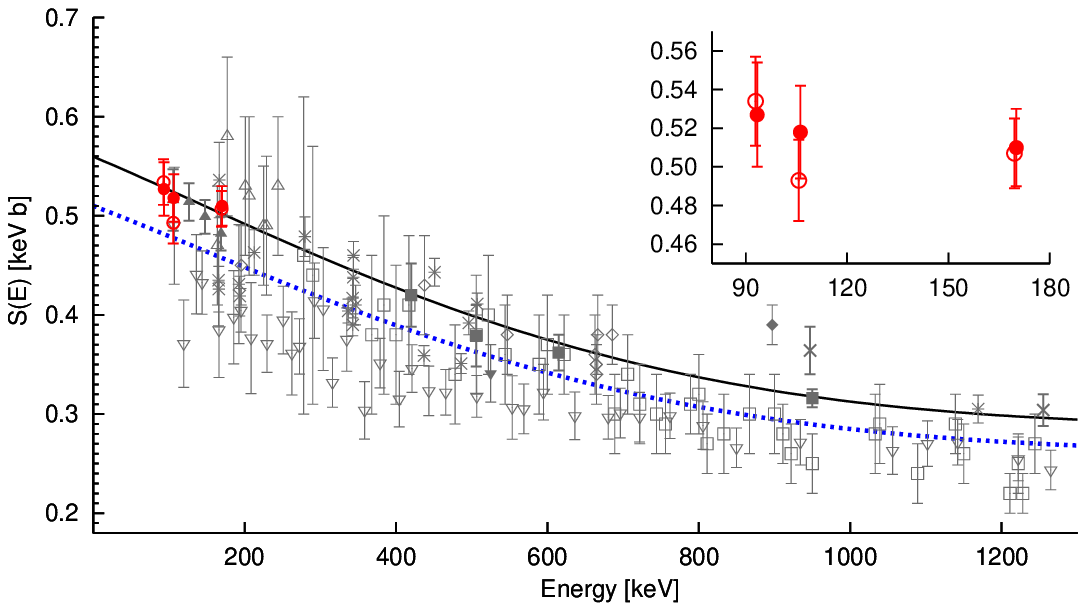}
    \caption{Overview of all available S-factor values for the $^3$He($\alpha$,$\gamma$)$^7$Be
    reaction.
Filled and open circles: present work. Prompt-$\gamma$ data: open
squares \cite{parker}, open triangles \cite{nagatani}, stars
\cite{osborne},open inverse triangles
    \cite{krawinkel}, filled inverse triangles \cite{alexander}, open diamonds
    \cite{hilgemeier}. Activation data: filled diamonds \cite{robertson},
     crosses \cite{osborne}, filled squares \cite {singh}, filled triangles \cite{bemmerer2,giuri}. Dashed line: most recent
    R-matrix fit \cite{descouvemont}, solid line: fit \cite{descouvemont} normalized to present data. In the inset a zoom of prompt $\gamma$ (filled circles)
    and activation (open circles) data obtained in this work.}
    \label{fattore}
\end{figure*}

 Results are reported in Table
\ref{tabella} and shown in Fig. \ref{fattore} together with all
previous literature data.
\begin{table*}[here!]
\begin{center}
\begin{ruledtabular}
\begin{tabular}{l c c c c c c c c }

E$\alpha$ & Method & E$_{eff}$ & $\sigma$(E$_{eff})$ &
S(E$_{eff}$) & $\Delta$S stat. & $\Delta$S syst. &
$\Delta$S red. syst.\\
(keV)     &    & (keV)     & (nbarn)             & (keV barn)    &  (keV barn)    & (keV barn)     &  (keV barn)\\
\hline
400 & p & 170.1 & 10.25 & 0.510 & 0.008 & 0.019 & 0.015 & \\
400 & a & 169.5 & 10.00 & 0.507 & 0.010 & 0.015 & 0.010 & \\
250 & p & 106.1 & 0.588 & 0.518 & 0.014 & 0.019 & 0.016 & \\
250 & a & 105.7 & 0.546 & 0.493 & 0.015 & 0.015 & 0.011 & \\
220 & p & 93.3 & 0.235 & 0.527 & 0.018 & 0.020 &  0.016 & \\
220 & a & 92.9 & 0.232& 0.534 & 0.016 & 0.017 &   0.013 & \\

\end{tabular}
\end{ruledtabular}
\caption{Cross section and S-factor results with corresponding
uncertainties for the prompt (p) and activation (a) methods. In
the last column the reduced systematic uncertainty is reported
where contributions common to the two methods have been
subtracted.} \label{tabella}
\end{center}
\end{table*}
For each couple of data (prompt $\gamma$ and activation) obtained
at the same E$_{\alpha}$, the effective energy (E$_{eff}$),
calculated as described in \cite{bemmerer2},
 is slightly different. Indeed the target of the prompt $\gamma$ experiment, defined by the inner
collimator (Fig. \ref{chamber}), is a fraction of the whole target
contributing to $^7$Be production. The E$_{eff}$ difference
corresponds to an S-factor change smaller than 0.1\% according to
the energy dependence given in \cite{descouvemont}.
 In the comparison
between prompt and activation S-factors,  we have therefore
neglected the E$_{eff}$ differences and considered a total
uncertainty given by the statistical and reduced systematic errors
summed in quadrature (Table \ref{tabella}). The mean percentage
difference between the S-factor values in Table \ref{tabella}
($\Delta$S=(S$_a$-S$_p$)/((S$_a$+S$_p$)/2)) is $\Delta$S$_{m}$=
-0.014$\pm$0.042. This result limits to +2.8\% (maximum
$\Delta$S$_{m}$ value at 1$\sigma$ level) possible non-radiative
contributions to the reaction cross section. S-factor activation
values at E$_{\alpha}$=400 and 250 keV are compatible with those
previously obtained at LUNA with the same setup
\cite{bemmerer2,giuri}. Considering the average of the new and old
activation values at the same beam energy, $\Delta$S$_{m}$ does
not change. A simultaneous measurement with both activation and
prompt $\gamma$ technique at energies around E$_{cm}$=1 MeV, where
the oldest activation experiments \cite{osborne,robertson} were
performed,  would be useful to look for non-radiative
contributions in a higher energy region than the one explored in
the present experiment.
 To deduce the extrapolated S(0),
the fit of \cite{descouvemont} has been rescaled using the present
activation and prompt $\gamma$ data separately. The weighted
average between the two S(0) values has been calculated adopting
as weights the statistical error obtained from the fit and the
reduced systematic error. We get S(0)=0.560$\pm$0.017 keV barn
where the final uncertainty also includes the systematic error
common to the two methods. Performing the same calculation
considering also the most recent and very accurate results from
\cite{singh,bemmerer2,giuri} or using the theoretical function
\cite{kajino2} adopted in the NACRE compilation \cite{nacre},
instead of the R-matrix fit by \cite{descouvemont}, the
extrapolated S(0) changes less than 1\%. Low energy accurate data
in fact minimize the uncertainty upon extrapolation. However, a
refined measurement of the slope of the S-factor in a wide energy
range would be useful to confirm theoretical calculations reducing
the uncertainty on the extrapolated S(0).
 The uncertainty on the predicted $^8$B neutrino flux due to
S$_{34}$ is now reduced from 7.5\% to 2.4\% and the total
uncertainty, including astrophysical parameters, goes from 12\% to
10\% \cite{garay}. Similarly, the uncertainty on $^7$Be predicted
flux goes from 9.4\% to 5.5\%, being the contribution of S$_{34}$
error reduced from 8\% to 2.5\% \cite{garay}.

\begin{acknowledgments}
A particular thank goes to Carlos Pe\~na Garay for fruitful and
enlightening discussions. The authors are indebted to the INFN
technical staff at Gran Sasso, Genova, Padova and Legnaro for
their support during the experiment. This work was supported by
INFN and in part by: TARI RII-CT-2004-506222, OTKA T42733 and
T49245, and BMBF (05Cl1PC1-1)
\end{acknowledgments}

\end{document}